\documentstyle[12pt,times,graphicx,subfigure,pstricks,amssymb]{article}
\newtheorem{theorem}{Theorem}
\newtheorem{meta-thm}[theorem]{Meta-Theorem}

\newtheorem{remark}[theorem]{Remark}
\newtheorem{definition}[theorem]{Definition}

\newcommand\beq[1]{ \begin{equation}\label{#1} }
\newcommand{\eeq}{ \end{equation} }
\newcommand{\beqno}{ \[ }
\newcommand{\eeqno}{ \] }
\newcommand\beqa[1]{ \begin{eqnarray} \label{#1}}
\newcommand{\eeqa}{ \end{eqnarray} }
\newcommand{\beqano}{ \begin{eqnarray*} }
\newcommand{\eeqano}{ \end{eqnarray*} }
\newcommand\equ[1]{{\rm (\ref{#1})}}

\def\real{\mathbb{R}}
\def\integer{{\mathbb Z}}
\def\torus{{\mathbb T}}

\begin{document}

\title{\bf Transient times, resonances and drifts of attractors
in dissipative rotational dynamics}

\author{ Alessandra Celletti\\
{\footnotesize Dipartimento di Matematica}\\
{\footnotesize Universit\`a di Roma Tor Vergata}\\
{\footnotesize Via della Ricerca Scientifica 1}\\
{\footnotesize I-00133 Roma (Italy)}\\
{\footnotesize \texttt{(celletti@mat.uniroma2.it)}}
\and
Christoph Lhotka\\
{\footnotesize Dipartimento di Matematica}\\
{\footnotesize Universit\`a di Roma Tor Vergata}\\
{\footnotesize Via della Ricerca Scientifica 1}\\
{\footnotesize I-00133 Roma (Italy)}\\
{\footnotesize \texttt{(lhotka@mat.uniroma2.it)}}
}

\maketitle

\begin{abstract}
In a dissipative system the time to reach an attractor is often influenced by
the peculiarities of the model and in particular by the strength of the
dissipation. In particular, as a dissipative model we consider the spin--orbit
problem providing the dynamics of a triaxial satellite orbiting around a
central planet and affected by tidal torques. The model is ruled by the
oblateness parameter of the satellite, the orbital eccentricity, the
dissipative parameter and the drift term.  We devise a method which provides a
reliable indication on the transient time which is needed to reach an attractor
in the spin--orbit model; the method is based on an analytical result, precisely
a suitable normal form construction. This method provides also information
about the frequency of motion. A variant of such normal form used to
parametrize invariant attractors provides a specific formula for the drift
parameter, which in turn yields a constraint - which might be of interest in
astronomical problems - between the oblateness of the satellite and its orbital
eccentricity.
\end{abstract}

\maketitle

\vglue.1cm
\noindent \bf Keywords. \rm Spin--orbit problem, transient time,
dissipative system, attractor.
\vglue.1cm

\section{Introduction}\label{sec:intro}
We consider a nearly--integrable dissipative system describing the motion of a
non--rigid satellite under the gravitational influence of a planet. The motion
of the satellite is assumed to be Keplerian; the spin--axis is perpendicular
to the orbit plane and it coincides with the axis whose moment of inertia is
maximum. The non--rigidity of the satellite induces a tidal torque provoking a
dissipation of the mechanical energy. The dissipation depends upon a
dissipative parameter and a drift term.  If the dissipation were absent, the
system becomes nearly--integrable with the perturbing parameter representing
the equatorial oblateness of the satellite. The overall model depends also on
the orbital eccentricity of the Keplerian ellipse. This problem is often known as
the \sl dissipative spin--orbit model \rm and it has been extensively studied
in the literature (see, e.g., \cite{Alebook}, \cite{CFL}, \cite{MDbook}).

The spin--orbit model exhibits different kinds of attractors, e.g. periodic,
quasi--periodic and strange attractors (compare with \cite{broer}, \cite{broersimo}, \cite{CIO}, \cite{Dor}). As it often happens in dissipative
system, the dynamics evolves in such a way that the attractor is reached after
an initial transient regime of motion. The prediction of the transient time to
reach the attractor is often quite difficult (see, e.g., \cite{hunt}, \cite{KS}), but it is obviously of pivotal
importance to test the reliability of the result (think, e.g., to the problem
of deciding about the convergence of the Lyapunov exponents). The first goal of
this paper is to give a recipe which allows to decide the length of the
transient time, namely the time needed to go over the transient regime and to
settle the system into its typical behavior.
Our study is based on the construction of a suitable normal form for
dissipative vector fields
(see \cite{CC12}, compare also with \cite{DGL}, \cite{fasso}, \cite{LI05}, \cite{pucacco}) that generalizes
Hamiltonian normal forms that are usually implemented around elliptic equilibria (see \cite{Ruz2011}).
We compute the frequency in the normalized variables and use
it - as well as its back--transformation to the original variables - for a
comparison with a numerical integration of the equations of motion. Several
experiments are performed as the strength of the dissipation is varied. It
should be kept in mind that in dissipative systems one has to tune the drift
parameter in order to get specific attractors, since it does not suffice to
modify the initial conditions like in the conservative case (\cite{CCL13},
\cite{CCARMA}). A different formulation of the normal form, precisely a
suitable parametric representation of invariant attractors, allows to obtain an
explicit form for the drift on the attractor. Taking advantage of the physical
definition of the drift term, precisely as a function of the eccentricity
(\cite{peale}, see also \cite{CL}), one can derive interesting conclusions on a
link between the oblateness parameter and the eccentricity associated to a
given invariant attractor. We believe that this constraint might be useful in
concrete astronomical applications.

\vskip.1in

This paper is organized as follows. In Section~\ref{spinorbit} we present the
equations of motion of the spin--orbit problem in the conservative and
dissipative cases. The construction of the normal form is developed in
Section~\ref{sec:nf}, while the parametric representation of invariant
attractors is provided in Section~\ref{sec:par}.  The investigation of the
transient time and the analysis of the drift term are performed in
Section~\ref{sec:app}.  Some conclusions are drawn in
Section~\ref{sec:conclusions}.

\section{The spin--orbit problem with tidal torque}\label{spinorbit}
In this Section we describe the spin--orbit model, providing the equation of
motion in the conservative case (Section~\ref{sosec}) and under the effect of a
tidal torque, due to the internal non--rigidity of the satellite
(Section~\ref{sodiss}).

\vskip.1in

\subsection{The conservative spin--orbit problem}\label{sosec}
The spin--orbit model describes the dynamics of a rigid body with mass $m$, say $\mathcal S$, that
we assume to have a triaxial structure with principal moments of inertia
$I_1\leq I_2\leq I_3$.
The satellite $\mathcal S$ moves under the gravitational
effect of a perturbing body $\mathcal P$ with mass $M$. Moreover, we make the following assumptions:

\begin{itemize}

\item[$i)$] the body $\mathcal S$ orbits on a Keplerian ellipse around $\mathcal P$; we denote by
$a$ and $e$ the corresponding semimajor axis and eccentricity;

\item[$ii)$] the rotation axis of $\mathcal S$ is assumed to coincide
with the direction of the largest principal axis of inertia;

\item[$iii)$] the spin--axis is assumed to be aligned with the orbit normal;

\item[$iv)$] all other perturbations, including dissipative effects, are neglected.

\end{itemize}

\begin{figure}[h]
\centering
\includegraphics[width=.5\linewidth]{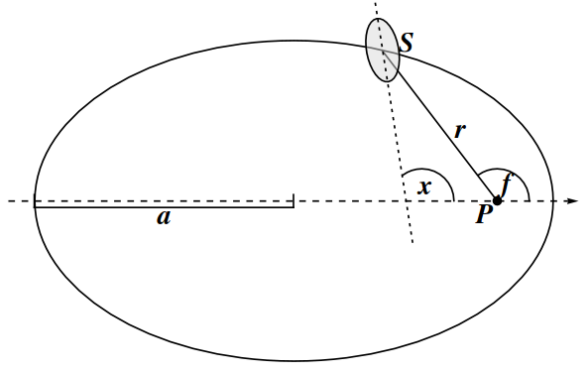}
\caption{The geometry of the spin--orbit problem: orbital radius $r$,
semi-major axis $a$, true anomaly $f$, rotation angle $x$.}
\label{so2}
\end{figure}

In order to simplify the notation, we normalize the units of measure;
precisely, the mean motion ${{{\mathcal G}M}\over a^3}$ (where ${\mathcal
G}$ is the gravitational constant) is normalized to one. An important role is played by
the following quantity, which is named the \sl equatorial ellipticity: \rm
\beqno
\varepsilon\equiv{3\over 2}{{I_2-I_1}\over I_3}\ .
\eeqno
To describe the rotation of $\mathcal S$ with respect to $\mathcal P$, we introduce the
angle $x$ spanned by the largest physical axis (that we assume to lie in the orbital plane)
with the perihelion line (see Figure~\ref{so2}).

The Hamiltonian function describing the spin--orbit model under the assumptions $i)$-$iv)$ is (see \cite{Alebook})
\beq{Hso}
{\mathcal H}(y,x,t)={{y^2}\over {2}}-{\varepsilon\over 2}({a\over r})^3
\cos(2x-2f)\ ,
\eeq
where $y$ is the momentum conjugated to $x$, $r$ is the orbital radius and $f$
is the true anomaly. Hamilton's equations associated to \equ{Hso} are given by
\beqano
\dot y&=&-\varepsilon ({a\over r})^3 \sin(2x-2f)\nonumber\\
\dot x&=&y\ ,
\eeqano
which are equivalent to the second--order differential equation
\beq{SO}
\ddot x+\varepsilon ({a\over r})^3 \sin(2x-2f)=0\ .
\eeq

\vskip.1in

\begin{remark}
\begin{itemize}

\item[a)]
The parameter $\varepsilon={3\over 2} {{I_2-I_1}\over I_3}$ plays the role of
the perturbing parameter: the Hamiltonian \equ{Hso} is integrable whenever
$\varepsilon=0$, which corresponds to the equatorial symmetry $I_1=I_2$.
For almost spherical bodies, like the Moon or Mercury, the value of
$\varepsilon$ is of the order of $10^{-4}$.

\item[b)]
It is important to stress that \equ{Hso} is a non--autonomous Hamiltonian function,
due to the fact that $r$ and $f$ are Keplerian functions of the time.
Introducing the eccentric anomaly $u$, defined in terms of the mean anomaly $\ell_0$ (which is a
linear function of time) through the well--known Kepler's equation $\ell_0=u-e\sin u$,
the orbital radius and the true anomaly can be determined by means of the following Keplerian expressions:
\beqa{rf}
r&=& a(1-e\cos u)\nonumber\\
f&=&2\arctan \Big(\sqrt{{1+e}\over{1-e}}\tan{u\over 2}\Big)\ .
\eeqa

\item[c)]
The Hamiltonian \equ{Hso} depends parametrically on the orbital eccentricity $e$ through $r$ and $f$
provided by \equ{rf}. We remark that in the case of circular orbits, equation \equ{SO} becomes integrable,
since $r$ is constant and $f$ coincides with time (up to a shift).
\end{itemize}
\end{remark}

\vskip.1in

Expanding $r$ and $f$ given in \equ{rf} in power series of $e$,
the Fourier expansion of equation \equ{SO} can be written as
\beqa{SO1}
\ddot x\ +\ \varepsilon\ \sum\limits_{m\not= 0,m=-\infty}^{+\infty}\
W({m\over 2},e)\ \sin(2x-mt)\ =\ 0\ ,
\eeqa
where we introduced the coefficients $W({m\over 2},e)$, decaying as powers of $e$
(see, e.g., \cite{Alebook}). Using a compact notation, we write \equ{SO1} as
\beq{comp1}
\ddot x+\varepsilon V_x(x,t)=0\ ,
\eeq
where $V=V(x,t)$ is a time--dependent periodic function (the subscript $x$ denotes
derivative with respect to the argument). In particular, we consider a trigonometric function by retaining in
\equ{comp1} just the most important harmonics (see \cite{Alebook}). Precisely, keeping the same notation $V$ for the
trigonometric approximation, we define
\beqa{Vtrig}
V(x,t)&\equiv&-\bigg[
\big(\frac{1}{2}-\frac{5}{4}e^2+\frac{13}{32}e^4\big)\cos(2x-2t)\nonumber \\
&&+\big(\frac{7}{4}e-\frac{123}{32}e^3\big)\cos(2x-3t)+
\big(\frac{17}{4}e^2-\frac{115}{12}e^4\big)\cos(2x-4t)
\bigg]\ .
\eeqa

We now introduce the definition of a $p:q$ \sl spin--orbit resonance \rm
for $p,q\in\integer$ with $q>0$ as a periodic solution of \equ{SO1},
say $t\in{\real}\to x=x(t)\in{\real}$, such that it satisfies
\beqno
x(t+2\pi q)=x(t)+2\pi p \qquad {\rm for\ any\ }\ t\in\real\ .
\eeqno
The above expression implies that the ratio between the period of revolution and the period
of rotation is equal to $p/q$. It is widely known that the Moon, alike most of the evolved
satellites of the Solar system, move in a 1:1 spin--orbit
resonance (usually referred to as the \sl synchronous \rm resonance);
within the Solar system only Mercury moves in a non--synchronous resonance (\cite{colombo}, \cite{CL2}), precisely in a 3:2
spin--orbit resonance\footnote{The astronomical consequence of a 1:1 resonance
is that the satellite always points the same face to the host planet.
Mercury's 3:2 spin--orbit resonance means that,
almost exactly, during two orbital revolutions around the Sun, Mercury makes
three rotations about its spin--axis.}, since twice the orbital period is equal to
thrice the rotational period within an error of the order of $10^{-4}$ (see
\cite{Alebook}).

\subsection{The dissipative spin--orbit problem}\label{sodiss}
Due to assumption $iv)$ of Section~\ref{sosec}, dissipative effects have been discarded and in particular
we neglected the effect of the non--rigidity of the satellite. This contribution, which turns out to be
the most relevant dissipative effect, induces a tidal torque (\cite{MD}, \cite{peale}), which can be written
as a function ${\mathcal T}$ depending linearly on the angular velocity $\dot x$:
\beq{T}
{\mathcal T}(\dot x;t)=-K_d\Big[L(e,t)\dot x-N(e,t)\Big]\ .
\eeq
In the above expression we have introduced the functions $L$ and $N$ as
$$
L(e,t)={a^6\over r^6}\ ,\qquad N(e,t)={a^6\over r^6}\dot f
$$
(recall that $r$ and $f$ are known functions of the time). Moreover, the coefficient
$K_d$ is the \sl dissipative constant, \rm whose explicit expression is given by
$$
K_d\equiv 3n\ {{k_2}\over {\xi Q}} ({R_e\over a})^3 {M\over m}\ ,
$$
where $n$ denotes the mean motion (that we have normalized to one),
$k_2$ is the so--called \sl Love number \rm (see \cite{GS}),
the constant $\xi$ is defined through $I_3=\xi mR_e^2$ with $R_e$
denoting the equatorial radius,
$Q$ is called the \sl quality factor \rm (providing the
frequency of oscillation with respect to the rate of dissipation of energy,
\cite{GS}). In order to compare the size of the dissipative effect with that of the conservative part,
we notice that astronomical measurements provide a value for $K_d$ of the order of $10^{-8}$ for
the Moon or Mercury.\\

In the following we reduce the tidal torque by considering (as in
\cite{CL}) its average over one orbital period. In particular, taking the average of
\equ{T} with respect to time one obtains (see \cite{peale})
\beq{24ter}
\bar {\mathcal T}\equiv \bar {\mathcal T}(\dot x)=-K_d\Big[\bar L(e)\dot x-\bar N(e)\Big]
\eeq
with
\beqa{LN}
\bar L(e)&\equiv& {1\over{(1-e^2)^{9/2}}}\ (1+3e^2+{3\over 8}e^4)\nonumber\\
\bar N(e)&\equiv& {1\over{(1-e^2)^6}}\ (1+{{15}\over 2}e^2+{{45}\over 8}e^4+
{5\over {16}}e^6)\ .
\eeqa
In conclusion, the following differential equation describes the
spin--orbit problem under the dissipative effect due to the tidal torque:
\beq{diss2}
\ddot x\ +\varepsilon\Big({a\over r}\Big)^3\ \sin(2x-2f)=-K_d
\Big[\bar L(e)\dot x-\bar N(e)\Big]\ .
\eeq
As in \equ{comp1}, we use a compact notation re-writing \equ{diss2} as
\beq{MD1}
\ddot x\ +\varepsilon V_x(x,t)=-\mu (\dot x-\eta)\ ,
\eeq
where we have introduced $\mu$ and $\eta$ as follows:
$\mu\equiv K_d \bar L(e)$, $\eta\equiv \bar N(e)/\bar L(e)$. As we can see,
$\mu$ depends on the dissipative constant (as well as on $e$), and therefore we call it \sl dissipative parameter, \rm
while $\eta$ is just a function of the eccentricity, and we call it
the \sl drift parameter\rm.

\begin{remark}\label{ennelle}
The tidal torque in \equ{MD1} vanishes for
$\dot x=\eta$; in view of \equ{24ter}, the tidal torque vanishes as far as
\beq{x}
\dot x\equiv {{\bar N(e)}\over {\bar L(e)}}=
{{1+{15\over 2}e^2+{45\over 8}e^4+{5\over {16}}e^6}\over
{(1-e^2)^{3\over 2}(1+3e^2+{3\over 8}e^4)}}\ .
\eeq
When $e=0$ equation \equ{x} implies that $\dot x=1$, which corresponds to the synchronous resonance.
For the actual Mercury's eccentricity amounting to $e=0.2056$, \equ{x} provides the value
$\dot x=1.256$, while for future use we notice that $e=0.285$ corresponds to $\dot x=1.5$, namely the 3:2
resonance.
\end{remark}

\section{A normal form construction}\label{sec:nf}
Our next task is to develop a normal form which transforms \equ{MD1} into a
system of equations which is normalized up to a given order (see \cite{CC12}).
This allows us to compute a normalized frequency, which will provide useful
information on the dynamical behavior of the model described by \equ{MD1}.

Let us write \equ{MD1} as the first--order differential system
\beqa{so1}
\dot x&=&y\nonumber\\
\dot y&=&-\varepsilon V_x(x,t)-\mu (y-\eta)\ .
\eeqa
Let us denote the frequency vector of motion associated to the
one--dimensional, time dependent equation
\equ{so1} as $\omega(y)=(\omega_0(y),1)$.
Assume that the vector field \equ{so1} is defined on a set $A\times \torus^2$,
where $A\subset \real$ is an open set.  Let $y_0\in A$ be an initial condition such
that the frequency $\omega_0=\omega_0(y_0)$ satisfies the following \sl
non--resonance condition: \rm
$$
|\omega_0 m+n|>0\qquad {\rm for\ any}\ \ (m,n)\in\integer^2\ ,\quad n\not=0\ .
$$
We look for a transformation of coordinates defined up to a suitable order
$N\in\integer_+$ in $\varepsilon,\mu$, say $\Xi_N:A\times {\torus}^2\rightarrow
{\real}\times{\torus}^2$, such that the new variables are $(Y,X)$ with
\beq{nf1}
(Y,X,t)=\Xi_N(y,x,t)\ ,\qquad Y\in{\real}\ ,\quad (X,t)\in{\torus}^2\ .
\eeq
In the transformed set of coordinates we require that the equations become:
\beqa{trasf}
\dot X &=& \Omega(Y;\varepsilon)+O_{N+1}(\varepsilon,\mu)\nonumber\\
\dot Y &=& -\mu(Y-\eta)+O_{N+1}(\varepsilon,\mu)\ ,
\eeqa
where $O_{N+1}(\varepsilon,\mu)$ denotes a function whose Taylor series
expansion in $\varepsilon$, $\mu$ contains only monomials $\varepsilon^j\mu^m$
with $j+m\geq N+1$.

According to \cite{CC12}, the transformation \equ{nf1} is obtained as the
composition of two transformations. The first one brings the original variables
$(x,y,t)$ into intermediate variables $(\tilde x,\tilde y,t)$, so to remove
terms depending on $\varepsilon$; then, from $(\tilde x,\tilde y,t)$ we
implement another change of variables to $(X,Y,t)$ in such a way to obtain
\equ{trasf}.

\vskip.1in

The normal form \equ{trasf} is particularly useful, since neglecting
$O_{N+1}(\varepsilon,\mu)$ one can integrate the second equation as
\beq{Y}
Y(t)=\eta+(Y_0-\eta)\ e^{-\mu(t-t_0)}\ ,
\eeq
where we denote by $(X_0,Y_0)$ the initial conditions at time $t=t_0$ in the
normal form variables.  The expression \equ{Y} shows that, in the approximation
obtained neglecting higher order terms, the solution tends to $Y=\eta$ as time
tends to infinity. Inserting \equ{Y} into the first of \equ{trasf} we obtain
the dependence of $\dot X$ on time, whose integration provides $X=X(t)$ with
$X(0)=X_0$. Indeed the solution \equ{Y} provides the \sl natural \rm attractor,
which can be found in the original coordinates by integrating equations
\equ{so1} with $\varepsilon=0$.

The local behavior near quasi--periodic attractors of some dissipative systems, precisely
\sl conformally symplectic \rm systems\footnote{A flow $f_t:{\mathcal M}\rightarrow{\mathcal M}$
defined on a symplectic manifold ${\mathcal M}$ is \sl conformally symplectic, \rm
when $f_t^*\Omega=e^{\mu t}\Omega$ for some $\mu$ real with $\Omega$ the symplectic form.
Notice that \equ{so1} is a conformally symplectic flow according to such definition (see \cite{CCL13}).},
has been studied in \cite{CCLlinear}. The main result of \cite{CCLlinear} is that there exists a
transformation of coordinates such that the time evolution becomes a rotation in the angles and a
contraction in the actions. The normal form \equ{trasf} is consistent with such result: indeed,
neglecting higher order terms, the expression \equ{Y} shows that the normalized action contracts
exponentially in the dissipative parameter, while the first of \equ{trasf} shows that the limiting
behavior of the normalized angle is a linear rotation with frequency $\Omega(\eta)$.

\vskip.1in

For details on the normal form algorithm used to obtain \equ{trasf} we refer to \cite{CC12}; here we just
state the final result.  At the normalization order $N=3$ the normalized
frequency $\Omega(Y;\varepsilon)$ in \equ{trasf} turns out to be
\beqa{norfre}
\Omega(Y;\varepsilon)&=&Y-\frac{\varepsilon^2}{8 (Y-1)^3}+\varepsilon^2\bigg[
\frac{1}{8} e^2 \left(\frac{5}{(Y-1)^3}+\frac{98}{(3-2 Y)^3}\right)+\\ \nonumber
&&\frac{1}{64} e^4 \left(-\frac{63}{(Y-1)^3}+\frac{3444}
{(2Y-3)^3}-\frac{578}{(Y-2)^3}\right)+\\ \nonumber
&&\frac{1}{768} e^6\left(\frac{31280}{(Y-2)^3}+\frac{390}{(Y-1)^3}+
\frac{45387}{(3-2Y)^3}\right)\bigg]\ ,
\eeqa
where we expanded the coefficients up to the 6th order in the eccentricity. The
expansion of the drift $\eta=\bar N(e)/\bar L(e)$ (with $\bar N$, $\bar L$ as
in \equ{LN}) to the same order is given by:
\beq{etaex}
\eta=1+6e^2+\frac{3 e^4}{8}+\frac{173 e^6}{8} \ .
\eeq
Equations \equ{trasf} together with \equ{norfre}, \equ{etaex} will be used in
Section~\ref{sec:norapp} for a dynamical investigation based on a normal form
approach; precisely, we will backtransform the frequency $\Omega$ in the
original variables and compare the result with a numerical integration.

\section{A parametric representation of invariant attractors}\label{sec:par}
With reference to equation \equ{MD1}, we introduce in this Section a parametric
representation of an invariant KAM attractor with Diophantine frequency; as it is
well known (\cite{CCL13}, \cite{CCARMA}), the equations for the embedding can
be solved under suitable compatibility conditions, which provide a relation
between the frequency and the drift. In particular, such compatibility
conditions allow us to provide an explicit computation of the drift that we
perform up to the 4-th order in the series development in the perturbing
parameter. These results are used in Section~\ref{sec:parapp} in order to
investigate in some specific cases the relation between the drift and the
frequency, as well as the dependence on the other parameters (most notably the
oblateness and the eccentricity).

Let us recall that the frequency vector of motion associated to \equ{MD1} is
written as $\omega=(\omega_0,1)$. We say that $\omega$ satisfies the
Diophantine condition, whenever the inequality \beq{DC} |\omega_0\ m+ n|^{-1}\
\leq\ C |m| \qquad {\rm for\ all}\ \ (m,n)\in {\mathbb Z}^2\ ,\quad m\not=0
\eeq
is satisfied for some positive real constant $C$. Next we provide the following
definition of a \sl KAM attractor \rm for \equ{MD1}.

\begin{definition}
A KAM attractor for \equ{MD1} with rotation number $\omega=(\omega_0,1)$
satisfying \equ{DC} is a two--dimensional invariant surface, described
parametrically by
\beq{par}
x=\theta+u(\theta,t)\ ,\quad (\theta,t)\in\torus^2\ ,
\eeq
where the flow in the parametric coordinate is linear, namely
$\dot\theta=\omega_0$, and where $u=u(\theta,t)$ is a suitable analytic,
periodic function, such that\footnote{The requirement \equ{diff} guarantees
that \equ{par} is a diffeomorphism.}
\beq{diff}
1+u_\theta(\theta,t)\not=0 \quad {\rm for\ all}\ (\theta,t)\in\torus^2\ .
\eeq
\end{definition}

Let $D$ be the partial derivative operator defined as
\beq{Dch7}
D\equiv \omega_0{\partial\over {\partial\theta}}+{\partial\over {\partial t}}\ .
\eeq
Inserting \equ{par} in \equ{MD1} and using the definition \equ{Dch7}, it is
readily seen that the function $u$ must satisfy the differential equation
\beq{D2}
D^2u(\theta,t)+\varepsilon V_x(\theta+u(\theta,t),t)=
-\mu\Big(\omega_0+Du(\theta,t)-\eta\Big)\ .
\eeq
Notice that the inversion of the operator $D^2$ provokes the appearance of the
well--known problem of the small divisors (\cite{Alebook}).  An approximate
solution of \equ{D2} can be found as follows. Let us expand $u$ and $\eta$ in
Taylor series of $\varepsilon$ as
\beq{series}
u(\theta,t)=\sum_{k=1}^\infty u_k(\theta,t)\varepsilon ^k\ ,\qquad
\eta=\sum_{k=0}^\infty \eta_k\varepsilon ^k\ .
\eeq
Inserting \equ{series} in \equ{D2} and equating same orders in $\varepsilon$,
one obtains the iterative equations
\beqa{iterative}
D^2 u_1(\theta,t)+\mu Du_1(\theta,t)&=&-V_x(\theta,t)+\mu\eta_1\nonumber\\
D^2 u_2(\theta,t)+\mu Du_2(\theta,t)&=&-V_{xx}(\theta,t)\ u_1(\theta,t)+
\mu\eta_2\nonumber\\
...\nonumber\\
D^2 u_k(\theta,t)+\mu Du_k(\theta,t)&=&S_k(\theta,t)+\mu\eta_k\ ,
\eeqa
where at the order $k$ the function $S_k$ is known and it depends on the
derivatives of $V$ as well as on $u_1$, ..., $u_{k-1}$. At each order one
determines first $\eta_k$ so that the right hand sides of \equ{iterative} have
zero average. After having determined $\eta_k$, the $k$--th equation in
\equ{iterative} can be used to find $u_k$ as the solution of the following
equation:
\beq{uk}
D^2 u_k(\theta,t)+\mu Du_k(\theta,t)=\widetilde S_k(\theta,t)\ ,
\eeq
where $\widetilde S_k$ has zero average (in fact, $\widetilde S_k=S_k-\bar
S_k$, where the bar denotes the average over $\theta$ and $t$).  To solve
\equ{uk}, let us expand $u_k$ in Fourier series as
$$
u_k(\theta,t)=\sum_{(m,n)\in\integer^2} \hat u_{mn}^{(k)}\ e^{i(m\theta+nt)}\ ,
$$
where $\hat u_{mn}^{(k)}$ denote the (unknown) Fourier coefficients of $u_k$.
Inserting the Fourier series in \equ{uk} and expanding also the left hand side
in Fourier series, we obtain
\beq{uS}
\sum_{(m,n)\in\integer^2} \Big[-(\omega_0 m+n)^2+i\mu(\omega_0 m+n)\Big]\ \hat u_{mn}^{(k)}\ e^{i(m\theta+nt)}=
\sum_{(m,n)\in {\mathcal I}_S^{(k)}} \hat S_{mn}^{(k)}\ e^{i(m\theta+nt)}\ ,
\eeq
where ${\mathcal I}_S^{(k)}$ denotes the set of the Fourier indexes of
$\widetilde S_k$. Equation \equ{uS} allows to determine $u_k$ as
\beq{uk1}
u_k(\theta,t)=\sum_{(m,n)\in {\mathcal I}_S^{(k)}} {{\hat S_{mn}^{(k)}}
\over {-(\omega_0 m+n)^2+i\mu(\omega_0 m+n)}}\ e^{i(m\theta+nt)}\ .
\eeq
Notice that the assumption \equ{DC} on $\omega_0$ guarantees that $u_k$ is well
defined (no zero divisors appear in \equ{uk1}).  An alternative (weaker)
assumption would be that $|\omega_0 m+n|>0$ for all $(m,n)\in{\mathcal
I}_S^{(k)}$.

Due to the fact that the function $V$ is assumed to be a trigonometric function
(see \equ{Vtrig}), also $S_k$ is trigonometric and it is convenient to write it
as
$$
S_k(\theta,t)=\sum_{(m,n)\in {\mathcal I}_S^{(k)}} \Big[\hat S_{mn}^{(k,c)}\cos(m\theta+nt)+\hat S_{mn}^{(k,s)}\sin(m\theta+nt)\Big]
$$
for suitable real coefficients $\hat S_{mn}^{(k,c)}$ and $\hat S_{mn}^{(k,s)}$.
Then in place of \equ{uk} we can write the solution in a real form which is
suitable for numerical computations as
\beqano
u_k(\theta,t)&=&-\sum_{m,n\in {\mathcal I}_S} \Big[ {{\hat S_{mn}^{(k,c)}}\over {(\omega_0 m+n)[(\omega_0 m+n)^2+\mu^2]}}\
\Big((\omega_0 m+n)\cos(m\theta+nt)-\mu\sin(m\theta+nt)\Big)\nonumber\\
&+&{{\hat S_{mn}^{(k,s)}}\over {(\omega_0 m+n)[(\omega_0 m+n)^2+\mu^2]}}\
\Big((\omega_0 m+n)\sin(m\theta+nt)+\mu\cos(m\theta+nt)\Big) \Big]\ .
\eeqano
In conclusion, the algorithm to compute the drift consists in solving
iteratively equations \equ{iterative} to obtain the functions $u_k$; at each
order, by imposing that the right hand sides have zero average, we obtain the
terms $\eta_k$ of the series expansion \equ{series} of the drift.

We provide here the $\eta_k$ expanded up to the third order (the fourth order
can be obtained through a reasonable computer time, but the expression becomes
too long to be displayed here):
\beqa{eta}
\eta_0&=&\omega_0\nonumber\\
\eta_1&=&\eta_3=0\nonumber\\
\eta_2&=&
-\frac{a^2}{2 (\omega-1) \left(\mu ^2+4 (\omega-1)^2\right)}
-\frac{b^2}{(2 \omega-3) \left(\mu ^2+(3-2 \omega)^2\right)} \nonumber \\
&-&\frac{c^2}{(2 \omega-1) \left(\mu ^2+4 (1-2 \omega)^2\right)}
\eeqa
with
\beqano
a=1 -\frac{5 e^2}{2} + \frac{13 e^4}{16}\ , \qquad
b=\frac{7 e}{2}-\frac{123 e^3}{16}\ , \qquad
c=-\frac{1}{6} e^2 \left(115 e^2-51\right)\ .
\eeqano
The above solution for $\eta$ defines the drift parameter in \equ{series} up to
finite order; this expression will be used in Section~\ref{sec:parapp} to
obtain in particular a constraint between the parameters of the model (precisely, the
oblateness parameter and the eccentricity).

\section{Transient time and attractor's drift}\label{sec:app}

We devote this Section to the discussion of some dynamical features of the
attractors associated to the model described by equation \equ{MD1}. In
particular, we provide a numerical method (see Section~\ref{sec:numapp}) which
allows to determine the time needed to reach an attractor. Indeed, the
computation of such transient time is a critical issue in the study of
dissipative systems. Next, we make a comparison between the results obtained
through the normal form computation of Section~\ref{sec:nf} as well as those
provided by the parametrization in Section~\ref{sec:par} with the results
obtained from direct numerical simulations (see Section~\ref{sec:norapp}). A
link between the oblateness parameter and the eccentricity is provided in
Section~\ref{sec:parapp} as a consequence of the construction of the drift term
through the parametric representation of Section~\ref{sec:par}.

\subsection{A numerical investigation of the transient time}\label{sec:numapp}
The numerical determination of the frequency of an attractor is usually a
cumbersome problem, since the time necessary to reach an attractor is often
strongly influenced by the strength of the dissipation.  With reference to
\equ{MD1}, if $\mu$ is small, then one typically needs to wait for a longer
time to be on the attractor, while for $\mu$ large, the transient time is
shorter. There is an intrinsic difficulty to measure the time to reach the
attractor and of course this might affect also the computation of the
frequency. In principle, Lyapunov exponents can be used to compute the
transient time by evaluating the moment at which they reach convergence.
It should be remarked that Lyapunov exponents provide more indications
on the dynamics, precisely on the rate growth of the phase space volume.
Notwithstanding these remarks, we present an alternative numerical recipe
which provides information on the attractor's frequency. It turns out, that
the computer execution time becomes longer with respect to the computation
based on our method (about a factor 2).

\begin{figure}
\centering
\includegraphics[width=0.4\linewidth]{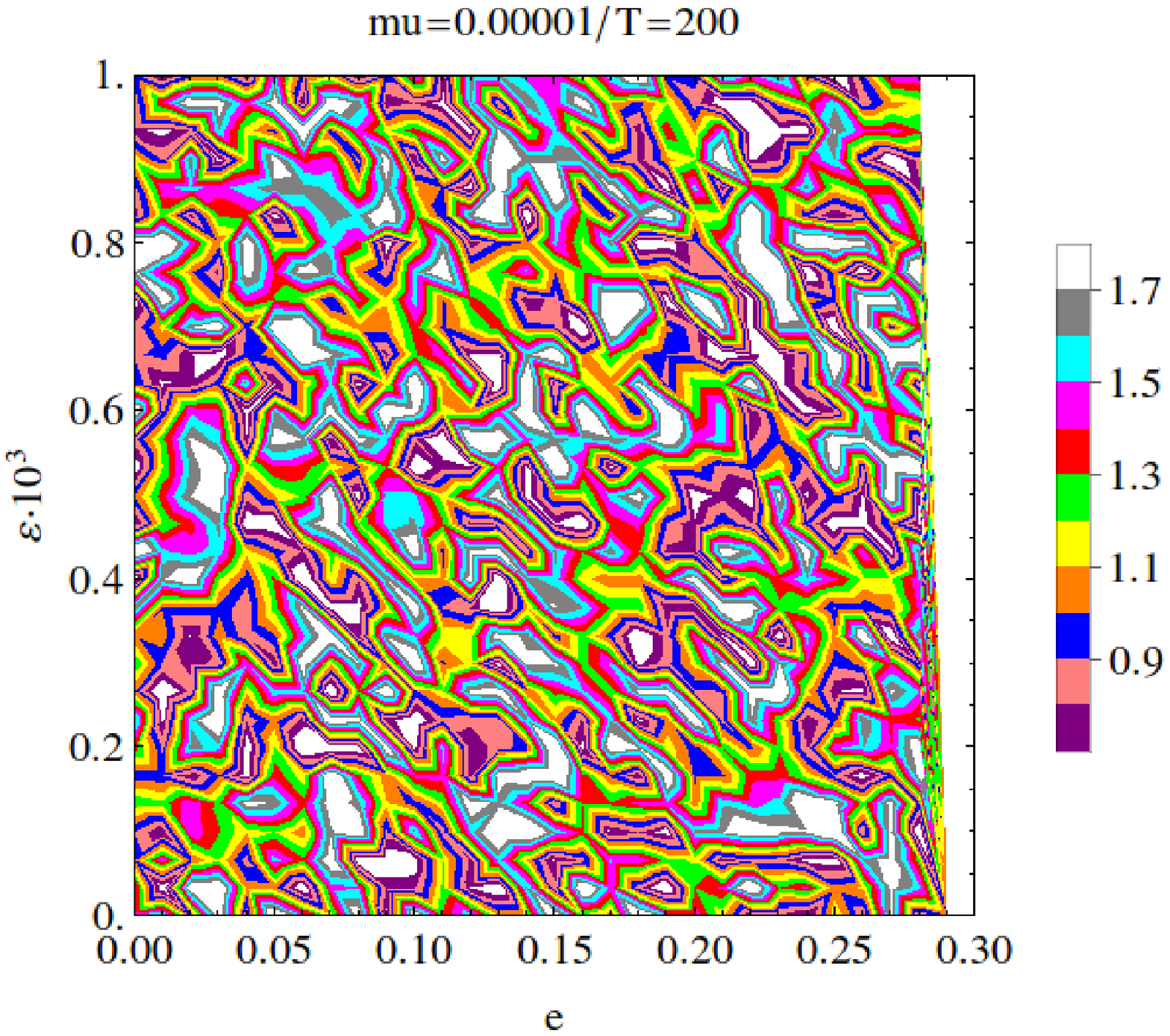}
\includegraphics[width=0.4\linewidth]{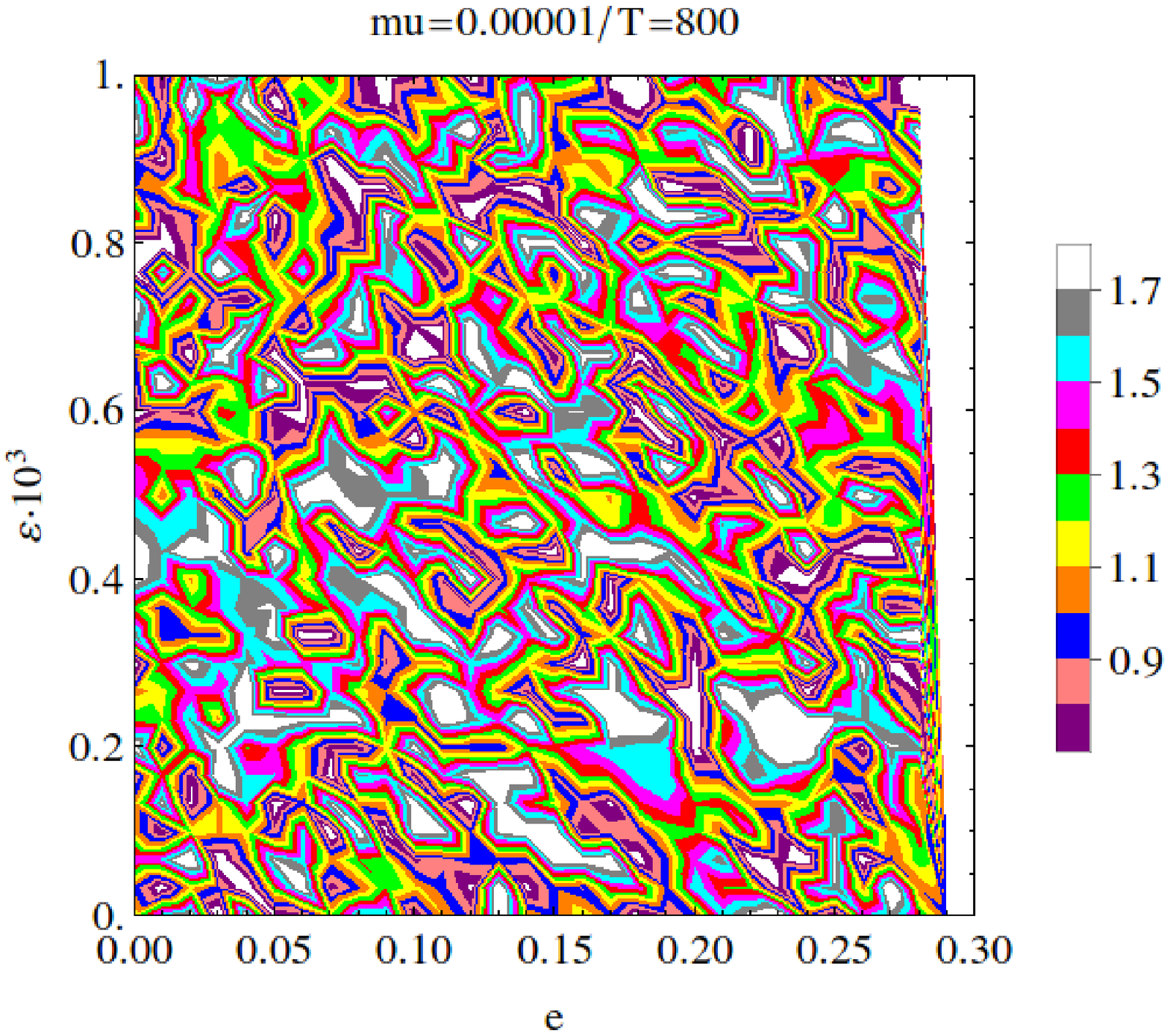}
\includegraphics[width=0.4\linewidth]{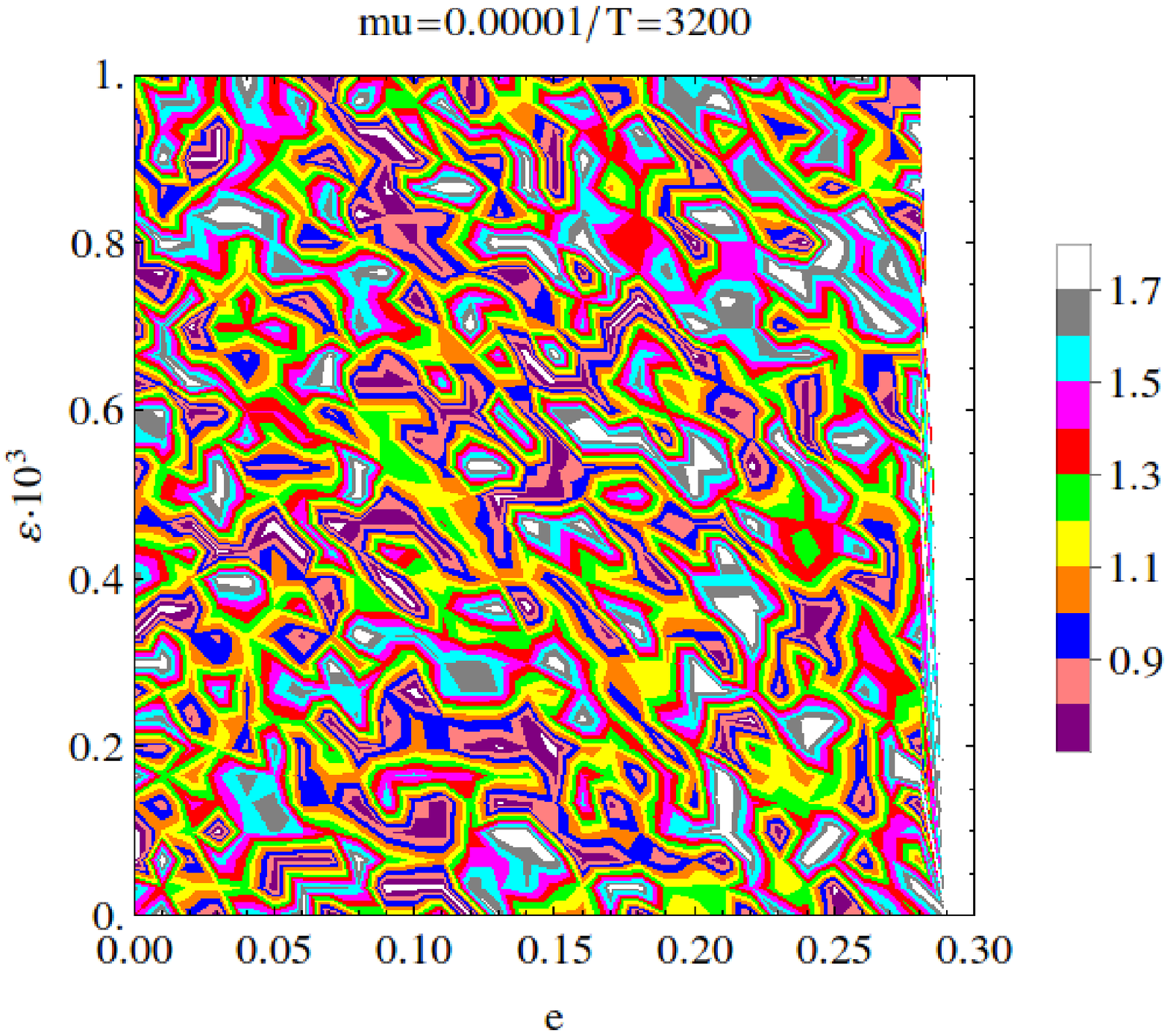}
\includegraphics[width=0.4\linewidth]{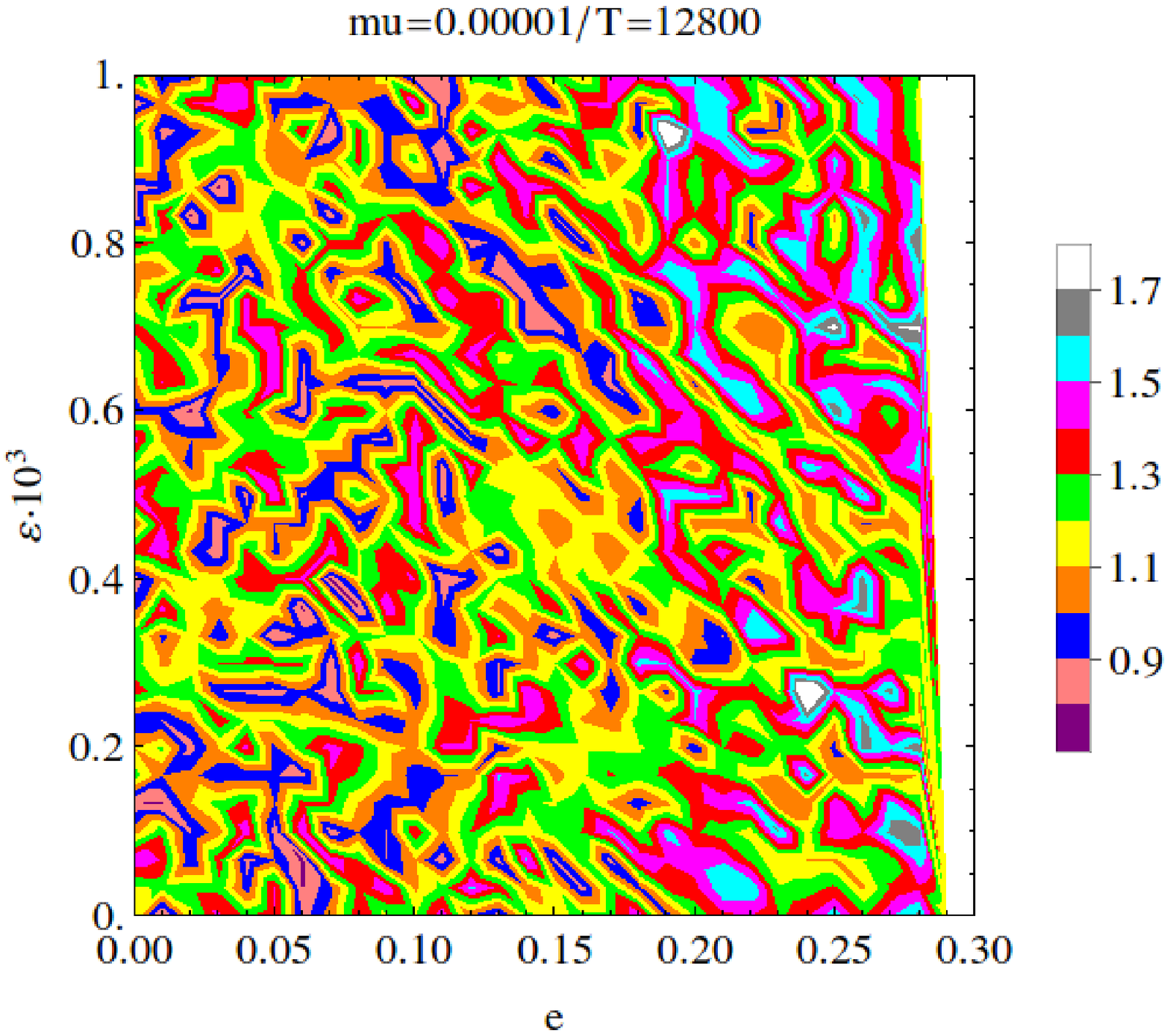}
\caption{The value $\omega_{num}$ in the space $(e,\varepsilon)$ for
$\mu=10^{-5}$ and integration times $T=200,800,3200,12800$.
The colors show the values of the frequency as given in the column bar. }
\label{f:num-5}
\end{figure}

\begin{figure}
\centering
\includegraphics[width=0.4\linewidth]{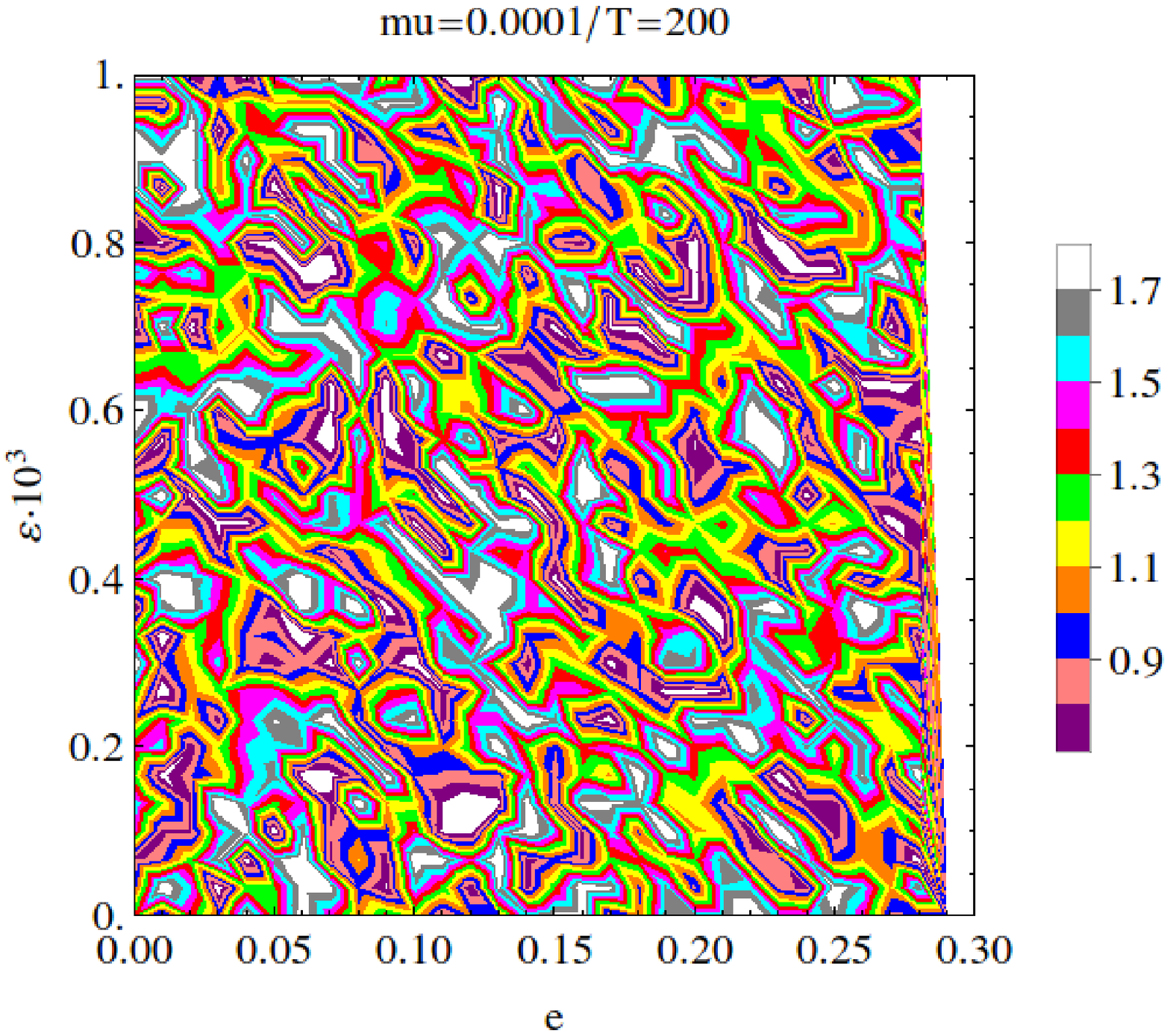}
\includegraphics[width=0.4\linewidth]{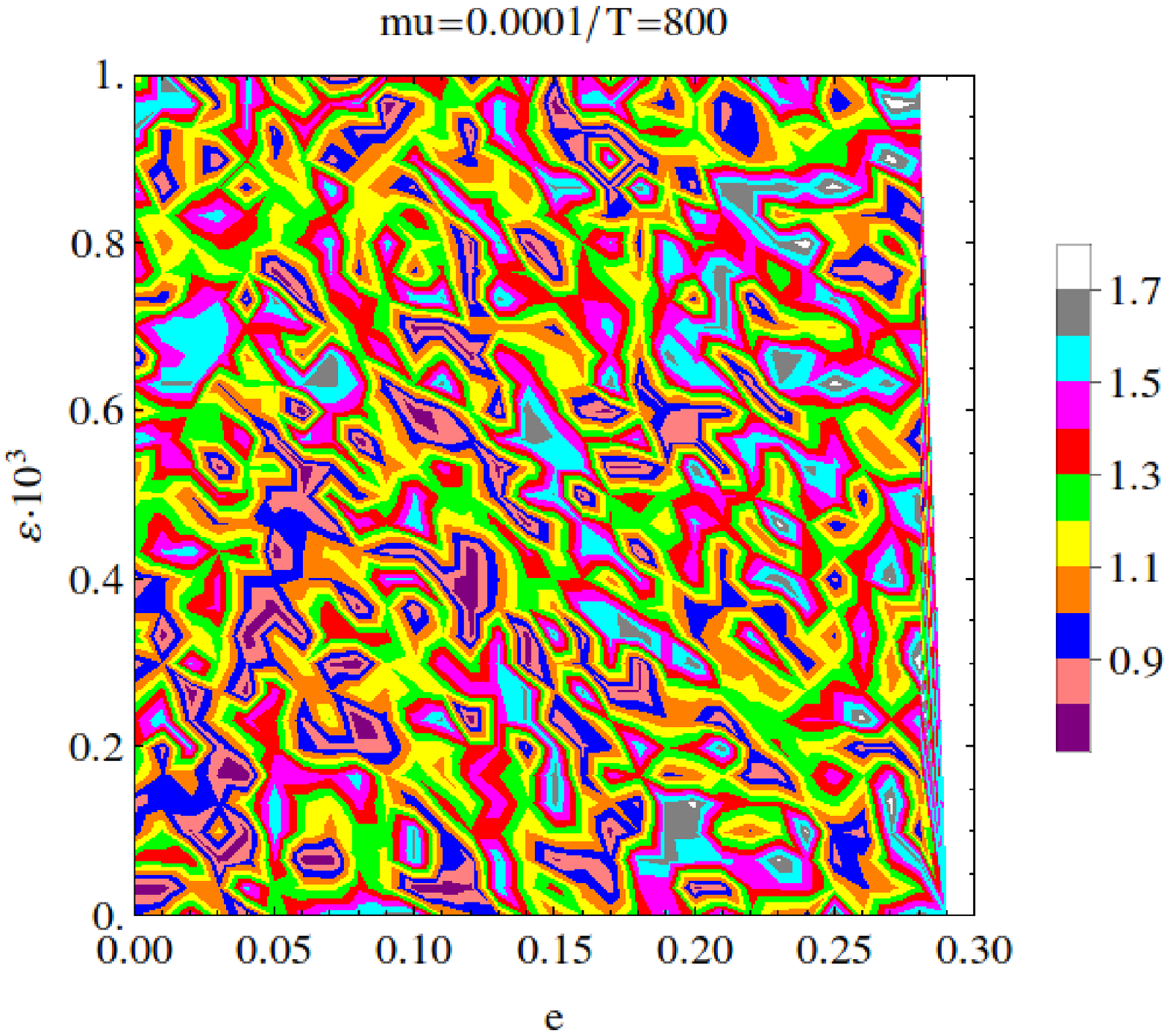}
\includegraphics[width=0.4\linewidth]{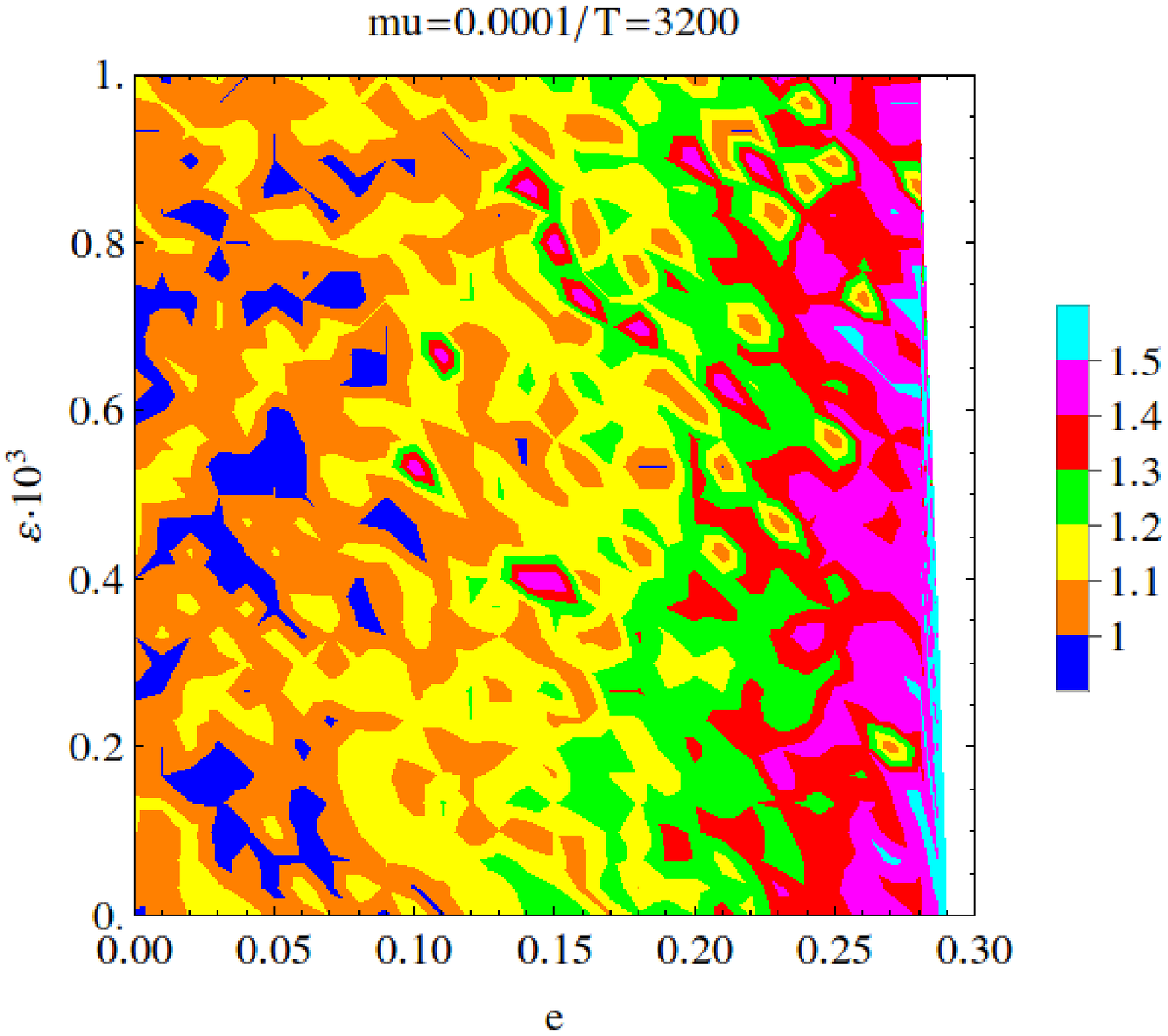}
\includegraphics[width=0.4\linewidth]{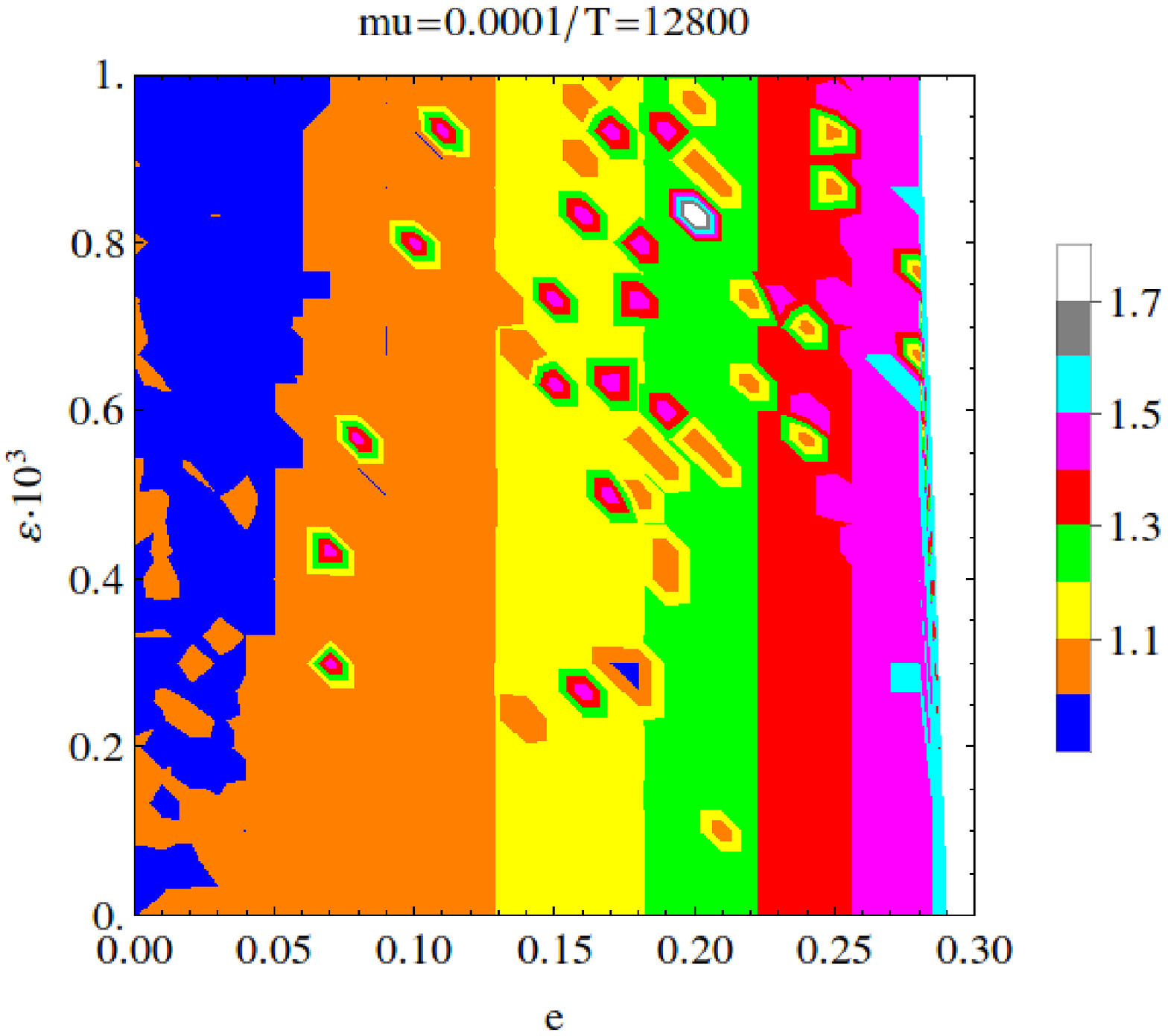}
\caption{The value $\omega_{num}$ in the space $(e,\varepsilon)$ for
$\mu=10^{-4}$ and different integration times $T=200,800,3200,12800$.
The colors show the values of the frequency as given in the column bar. }
\label{f:num-4}
\end{figure}

\begin{figure}
\centering
\includegraphics[width=0.4\linewidth]{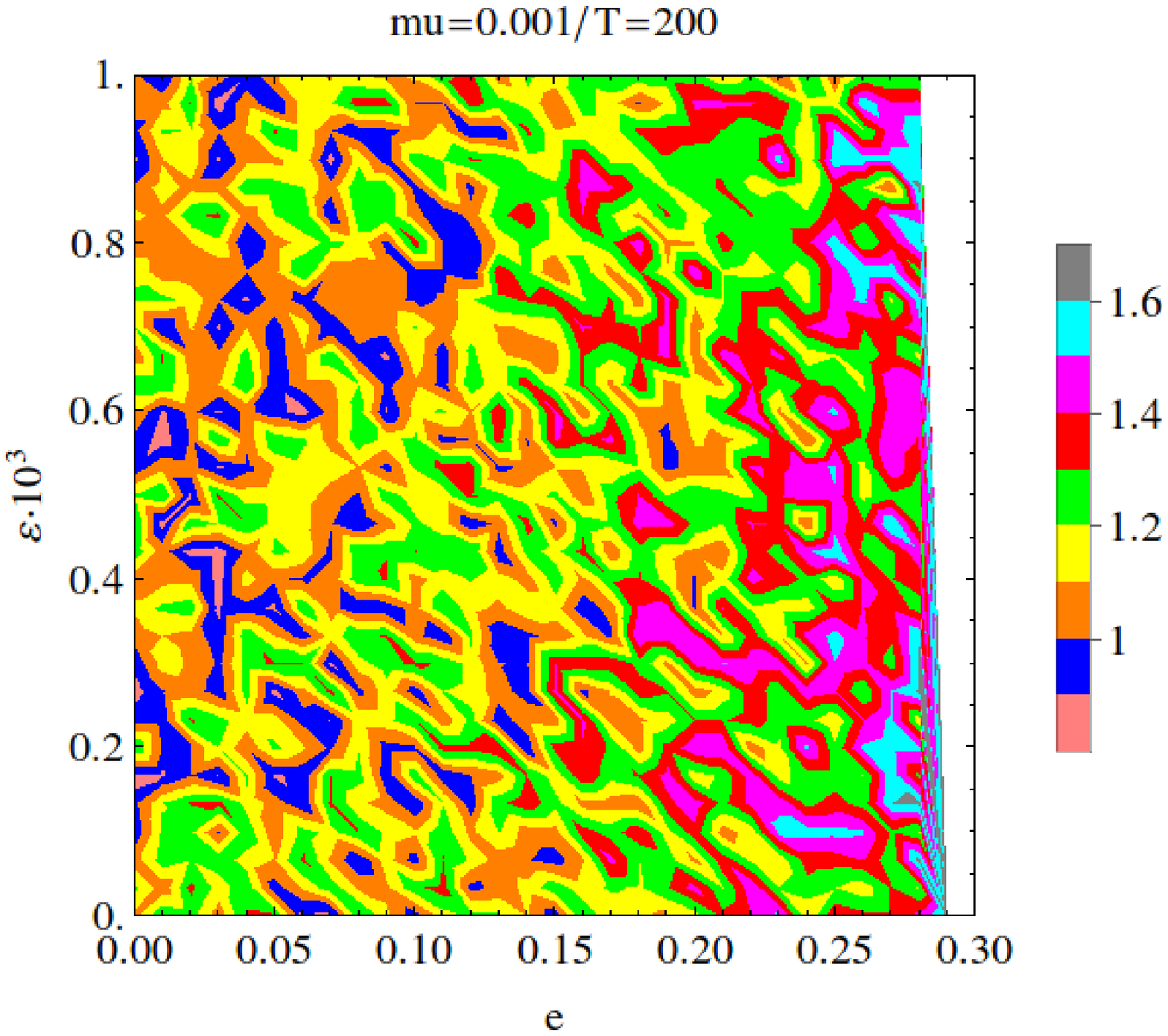}
\includegraphics[width=0.4\linewidth]{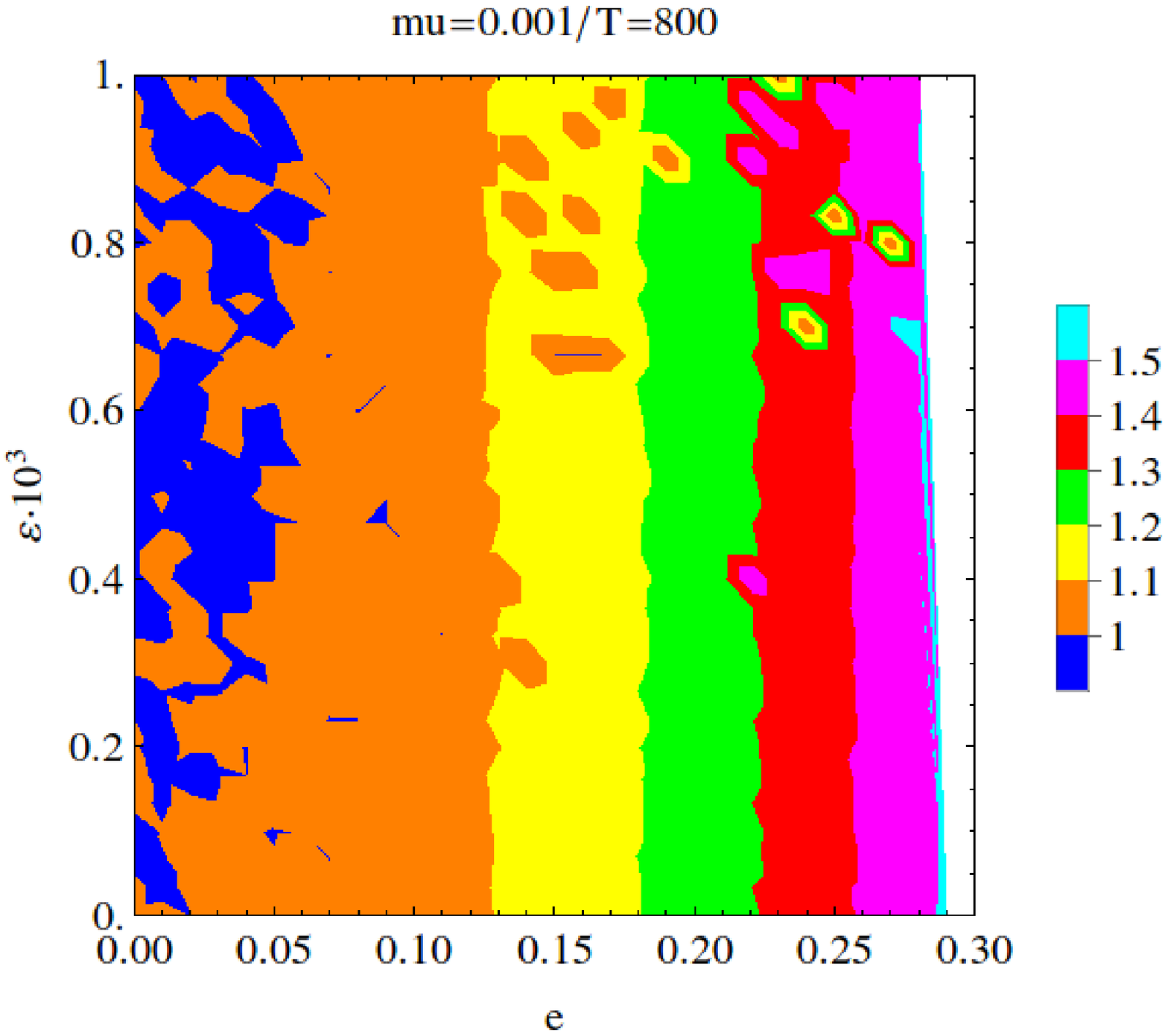}
\includegraphics[width=0.4\linewidth]{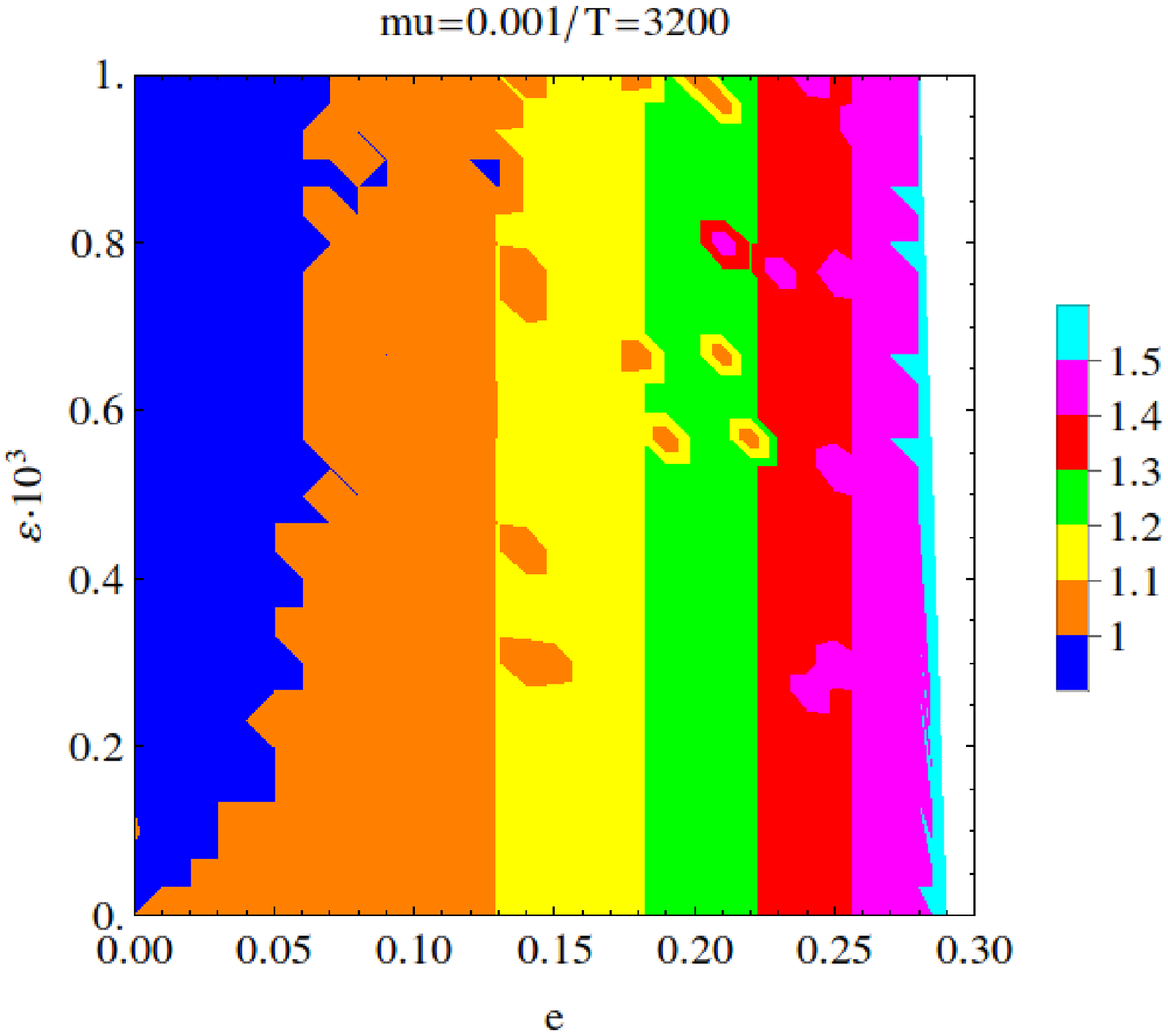}
\includegraphics[width=0.4\linewidth]{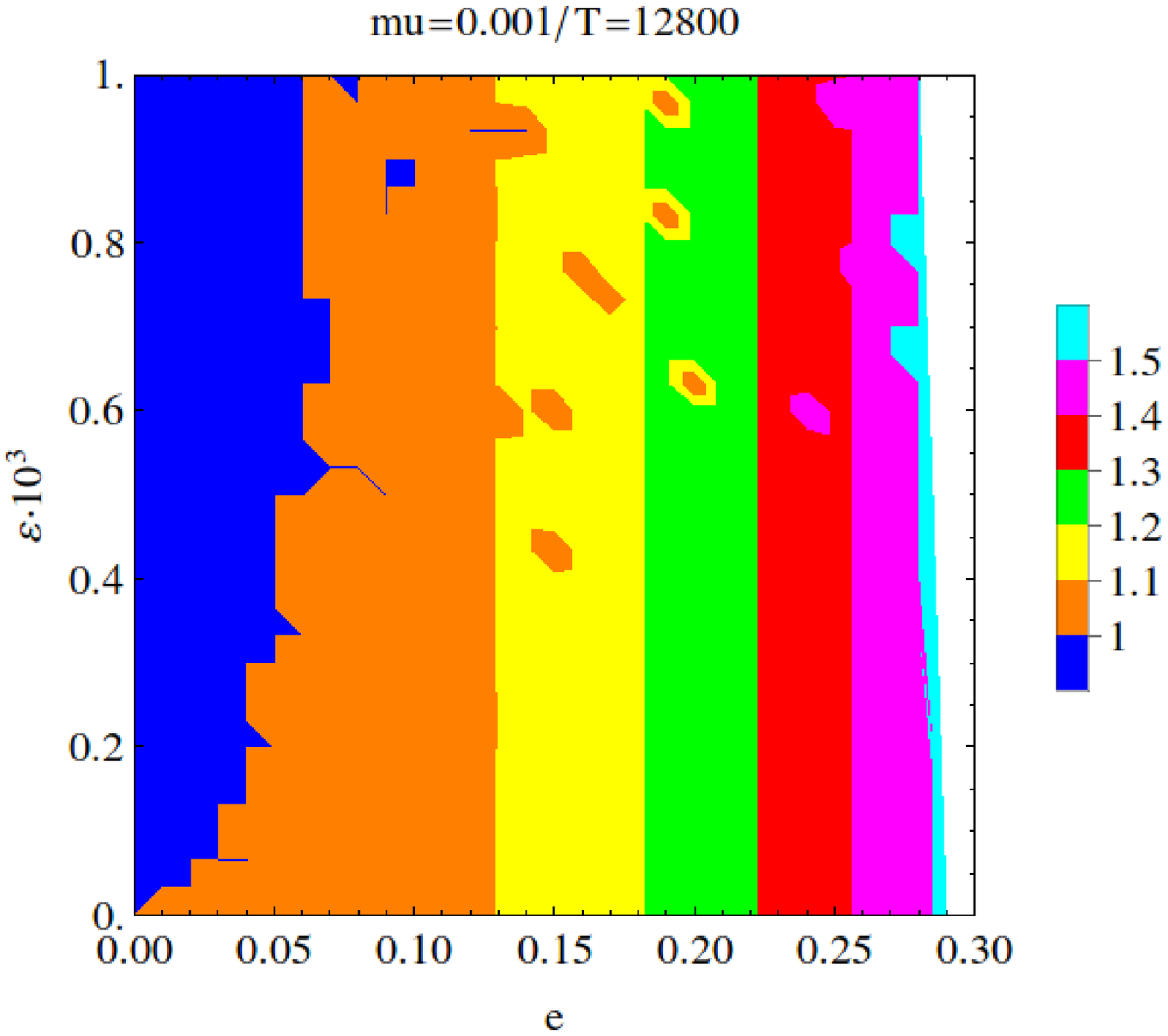}
\caption{The value $\omega_{num}$ in the space $(e,\varepsilon)$ for
$\mu=10^{-3}$ and different integration times $T=200,800,3200,12800$.
The colors show the values of the frequency as given in the column bar. }
\label{f:num-3}
\end{figure}

First, we integrate the original equations of motion \equ{so1} with $V$ given
in \equ{Vtrig} in the parameter space \(\mu \times \varepsilon \times e\) for
fixed \(\mu\) (e.g. $\mu=10^{-3},10^{-4},10^{-5}$), on a finite grid (e.g.
\(30\times 30\) values) in given intervals of $\varepsilon$, $e$ (e.g. \(0\leq
\varepsilon \leq 10^{-3}\), \(0\leq e\leq 0.3\)). The value of the drift has
been computed as a function of the eccentricity as $\eta=\bar N(e)/\bar L(e)$.
We use different integration times \(t_T=2\pi  T\) with
\(T=100,200,400,\ldots,12800\) (the last value is taken in order to obtain
results within a reasonable computer time).

Our choice of the parameters is made to cover the regime of interest for physical
applications. The upper bound 0.3 on the eccentricity encompasses both the eccentricity
of Mercury, about equal to 0.2, which is the largest body in a spin--orbit resonance,
and it is bigger than the value 0.285 which corresponds to the 3:2 resonance (see
Remark~\ref{ennelle}). Moreover, typical values for $\varepsilon$ are $10^{-4}$ for the
Moon or Mercury, $10^{-3}$ or even larger for the irregular satellites of our solar system.
Using $e\sim0.3$, $\varepsilon\sim\mu\sim10^{-3}$, a third order normal form gives an error,
e.g. in (17), of about $10^{-12}$, which is well below the numerical error that
we introduced in our numerical studies.

Let $\{x_k,y_k\}$ be an orbit computed at times multiple of $2\pi$, namely
\(x_k=x(2\pi  k)\), \(y_k=y(2\pi  k)\) with \(k=1,\ldots ,T\). We are
interested in the limiting behavior of the frequency \(\omega _k=\omega
\left(y_k\right)\) as \(k\) approaches \(T\). Let us focus on the value of the
frequency when approaching the maximum time $T$, say \(k=9/10 T,\ldots ,T\).
From the expression $$ \omega_{num}=\frac{10}{T}\sum _{k=9/10T}^T \omega_k\ ,
$$ we are able to estimate the mean value of \(\left\{\omega_k\right\}\) in the
proximity of $T$, taking the first part of the orbit as transient.  Next step
consists in finding a linear model of the form \(\omega_k =\omega_{num}+\sigma
k\) through \(\left\{\omega_k\right\}\): the parameter \(\sigma\) quantifies
the slope in \(\left\{\omega_k\right\}\).  We claim that \(\omega_{num}\) is
close to the final frequency on the attractor if \(\sigma \to 0\). We provide
now an example and we add a discussion on $\sigma$ at the end of this Section.

\vskip.1in

\begin{figure}
\centering
\includegraphics[width=0.55\linewidth]{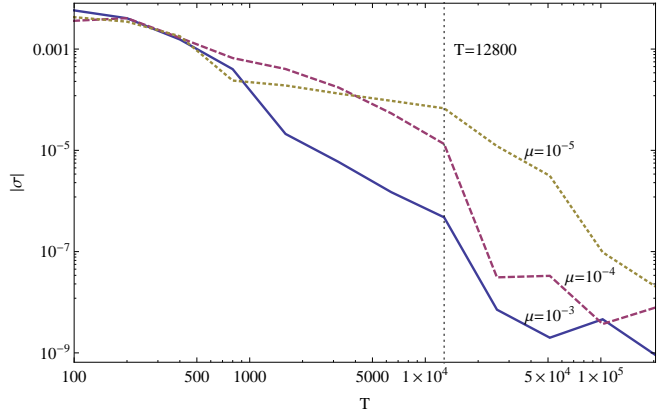}
\caption{The maximal absolute value of $\sigma$ versus the integration time
$T$ for different values of the dissipative parameter $\mu$. The larger $\mu$
or the longer the integration time $T$, the smaller the linear drift $|\sigma|$
(we increased $T$ for these specific cases to reach the error of the numerical
integration scheme, that also explains the fluctuations of $\sigma$ close to
$10^{-8}$).}
\label{fig:sig}
\end{figure}

We implement this procedure in a specific case, which is illustrated in
Figure~\ref{f:num-5}.  We start with a sample where \(\omega_{num}\) does not
coincide with the final frequency on the attractor. In fact, for \(\mu
=10^{-5}\) no regular pattern for \(\omega_{num}\) can be seen in
Figure~\ref{f:num-5}. The dynamics takes place in an intermediate regime, where
\(\omega_{num}\) still suffers huge variations in time. In this example, the
slope \(\sigma\) is still of the order of \(10^{-3}\) for \(T=100\) and
\(10^{-5}\) for \(T=12800\). After an integration time long enough, at
\(T=12800\) we start observing some structures, since \(\omega_{num}=1\)
accumulates for small \(e\), while \(\omega_{num}=1.5\) is obtained for larger
values of the eccentricity. This behavior would be more evident for longer
integration times and it will occur for a larger value of the dissipative
parameter (see Figures~\ref{f:num-4} and \ref{f:num-3}). The conclusion is that
for this set of parameters a longer time is necessary to obtain that the
dynamics has reached the attractor.  If we increase the dissipative parameter,
we obtain better results on the above time scales. In fact, for \(\mu
=10^{-4}\) we find (see Figures~\ref{f:num-4} and \ref{fig:sig}) that \(\sigma\) runs from
\(10^{-3}\) for \(T=100\) to \(10^{-6}\) for \(T=12800\). Therefore, increasing
\(\mu\) by one order of magnitude the slope \(\sigma\) decreases by one order
at \(T=12800\). For integration times long enough, we also clearly see that we
get a frequency $\omega_{num}$ which is almost constant for a fixed value of
the eccentricity as $\varepsilon$ varies.

When we increase further the strength of the dissipation, we see that the
attractor is reached on a shorter time scale.  Indeed, for \(\mu =10^{-3}\) the
slope \(\sigma\) ranges from \(10^{-3}\) for \(T=100\) to \(10^{-7}\)  for
\(T=12800\). Again we see a very clear separation of \(\omega_{num}\) at fixed
values of \(e\), and that the dependency of \(\omega_{num}\) on \(\varepsilon\)
is small (see Figure~\ref{f:num-3}).

It is possible to quantify the level of accuracy of our numerical approach by
investigating the size of $\sigma$ and its dependency on the integration time.
In Figure~\ref{fig:sig} we provide the absolute value of $\sigma$ versus the
integration time $T$ that we increased up to $T=102 400$ to be able to see
also the limiting behaviour for all values of the parameters, including $\mu=10^{-5}$.
We clearly see that the larger
the value of $\mu$ or the longer the integration times $T$, the smaller the
value of $\sigma$.  Such behavior of $\sigma$ provides a strong indication
on the transient time to reach the attractor.

\vskip.1in

\subsection{Using the normal form approach}\label{sec:norapp}

We proceed to compute the normalized frequency, that we can obtain by
implementing the normal form described in Section~\ref{sec:nf}. Precisely, we
use the solution for the normalized frequency \equ{norfre}, where (with a
little abuse) we replace $Y$ by its limiting value $Y_\infty=\eta$. The drift
$\eta$ is expanded up to finite order in the eccentricity as in \equ{etaex}.
This yields a good approximation of the normalized frequency that we call
$\Omega_{app}=\Omega_{app}(\varepsilon,e)$ .  The analytical result will be
compared with the frequency $\omega_{num}$, that we obtained from the numerical
simulations described in Section~\ref{sec:numapp}.

\begin{figure}
\centering
\includegraphics[width=0.45\linewidth]{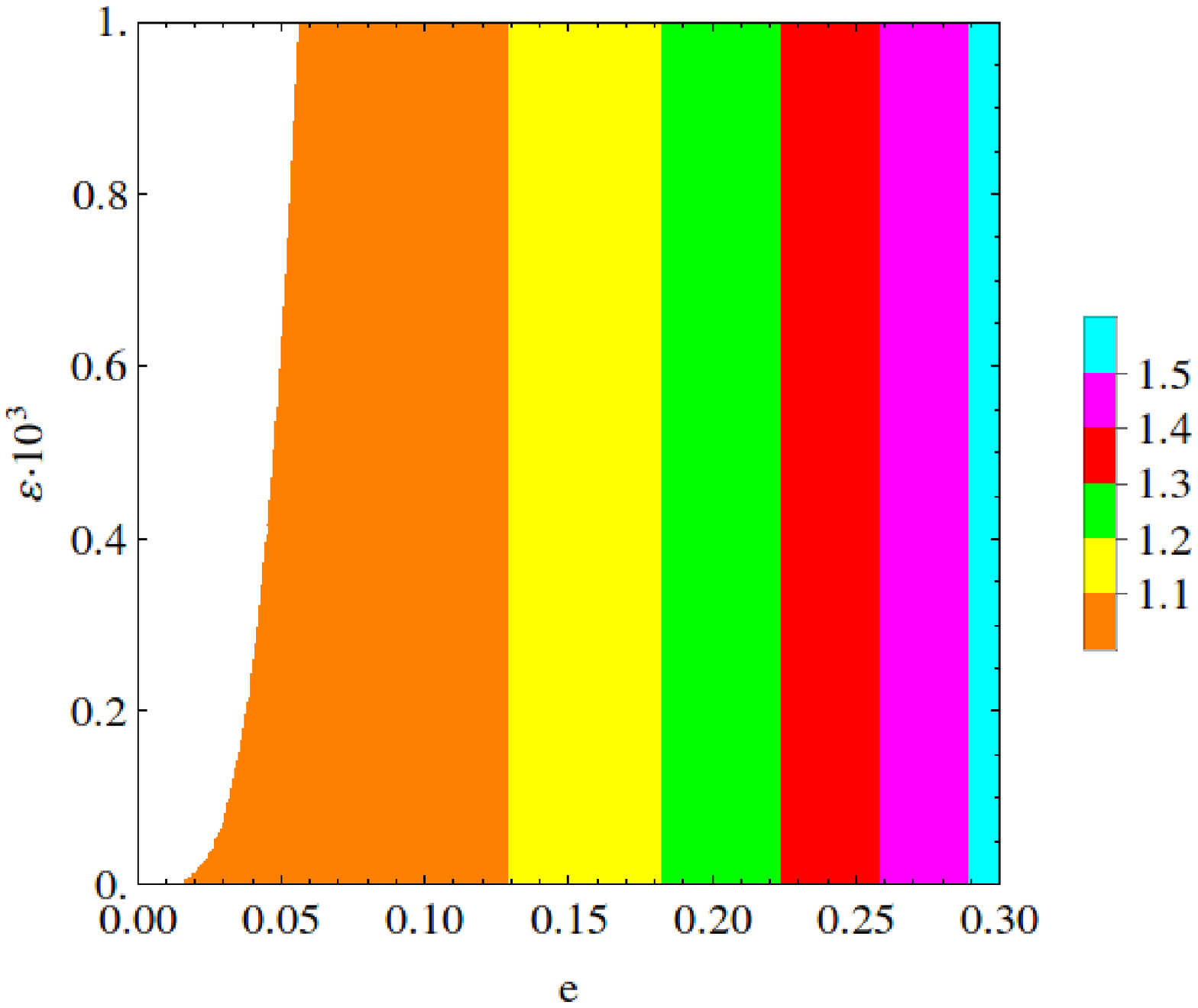}
\includegraphics[width=0.45\linewidth]{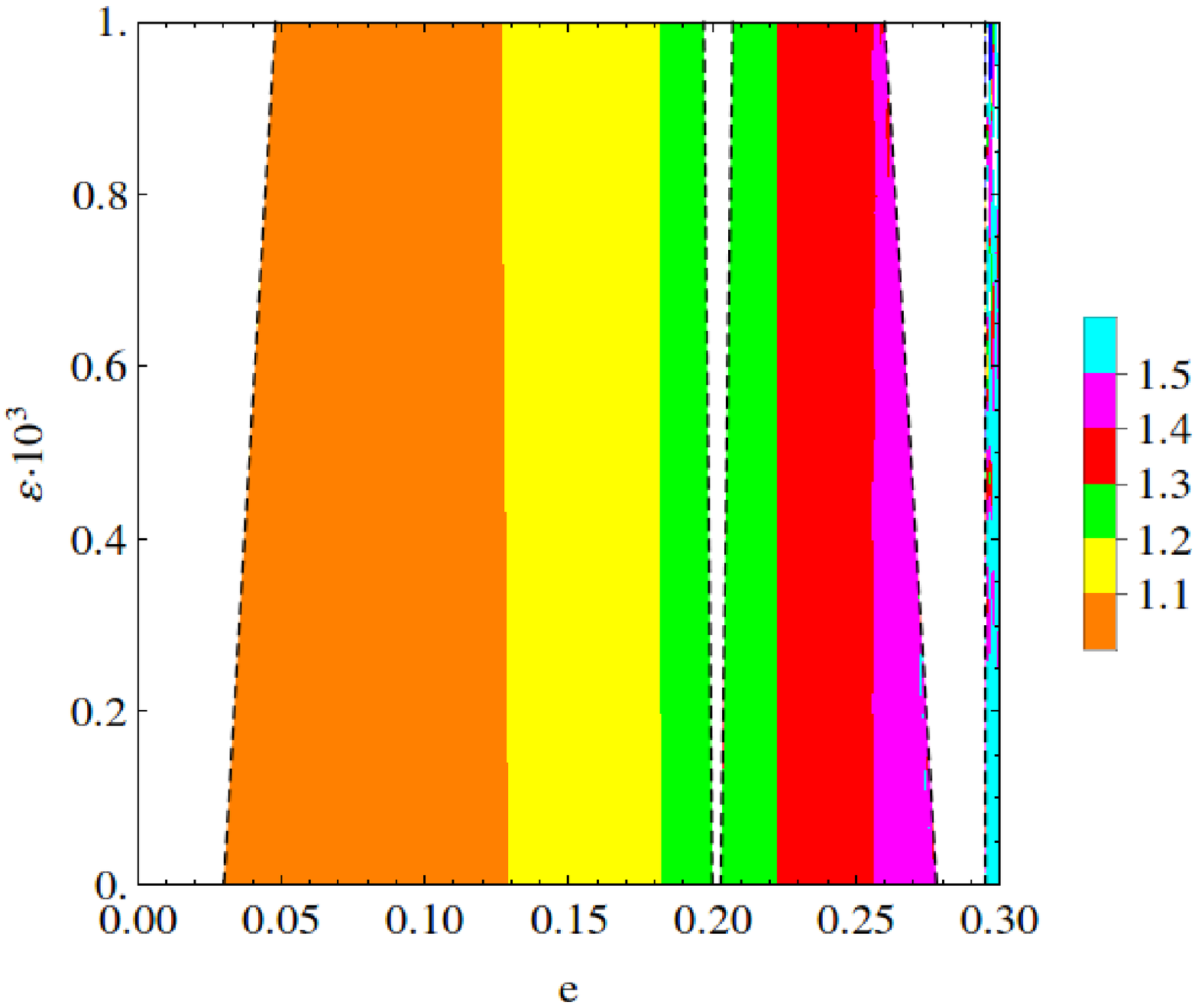}
\caption{Frequencies $\Omega_{app}$ and
\(\omega_{app} (y(Y_\infty,X_\infty;\varepsilon,\mu))\) in the
parameter space $(e,\varepsilon)$
obtained from the normal form approach: normalized frequency (left),
back-transformed frequency (right) with $\mu=10^{-3}$.
The figures have to be compared with the panels of Figure~\ref{f:num-3}.}
\label{f:nor}
\end{figure}

\vskip.1in

In Figure~\ref{f:nor} (left) we show the frequency \(\Omega_{app}\) obtained
from the normal form equation in the plane \((\varepsilon, e)\).  By comparing
Figure~\ref{f:nor} with Figures~\ref{f:num-5}--~\ref{f:num-3}, one can see that
the frequencies calculated using the normal form coordinates are in good
agreement with the frequencies that we obtain from the numerical approach.  We
also report the frequency that we calculated in the normalized variables in
terms of the original variables. To this end we need to compute the inverse of
the transformation \(\Xi_N(y,x,t;\varepsilon,\mu)\) in \equ{nf1}, say
$\Xi_N^{-1}(Y,X,t;\varepsilon,\mu)$, replace for \(X,Y\) the limits \(X_{\infty
},Y_{\infty }\), and calculate the average over time to get
\(\omega_{app}(y(Y_\infty,X_\infty;\varepsilon,\mu))\).  The results concerning
the normalized frequency in the original variables are given in
Figure~\ref{f:nor} (right); the values of Figure~\ref{f:nor} (right) and those
of Figure~\ref{f:num-3} (last panel) perfectly agree in the non-resonant
regime, while close to the main resonances corresponding to the white empty
zones (precisely, 1:1, 5:4, 3:2 resonances, corresponding to the outermost left
zone close to $e=0$, the tongue originating at $e=0.2$ and the right white zone
at about $e=0.29$), the non-resonant normal form together with the
transformation cannot be used, due to the accumulation of zero divisors in the
denominators of the analytic expansions. Note that in Figure~\ref{f:nor} (left)
the plot is independent of $\mu$, since
$\Omega_{app}=\Omega_{app}(\varepsilon,e)$, see \equ{norfre}, turns out to be
independent of $\mu$, while there is a parametric dependency of $\omega_{app}$
on $\mu$ due to the fact that we used the inverse transformation
$\Xi_N^{-1}(X,Y,t;\varepsilon,\mu)$.

\vskip.1in

\subsection{A constraint resulting from the parametrization}\label{sec:parapp}
In the previous sections we concentrated on the computation of the frequency of
motion for given values of the parameters, including the drift parameter.
Nevertheless, in many physical situations we might be interested to focus on a
specific frequency; however, fixing the frequency means that we need to find
the drift parameter in terms of such frequency as well as $\varepsilon$, $e$.
This can be done in the case of non--resonant frequencies by implementing the
parametric representation of invariant tori described in Section~\ref{sec:par}.
Indeed, in order to be able to solve the equations \equ{iterative} we must
compute the terms $\eta_k$ of the series expansion of $\eta$ in \equ{series} as
given by the expressions \equ{eta}, say $\eta=\eta(\omega,e,\varepsilon,\mu)$.
Normally the procedure is to fix $\varepsilon, \mu$ and to find a relation
between the drift $\eta$ and the frequency. Given that $\eta$ is a function
of the eccentricity, this procedure amounts to producing a relation between
the eccentricity and the frequency for fixed values of $\eta,\mu$. However,
we can decide to fix the frequency $\omega$ and to let $\varepsilon$ vary
in order to obtain an expression between the eccentricity $e$ and the shape
parameter $\varepsilon$.

\begin{figure}
\centering
\includegraphics[width=0.65\linewidth]{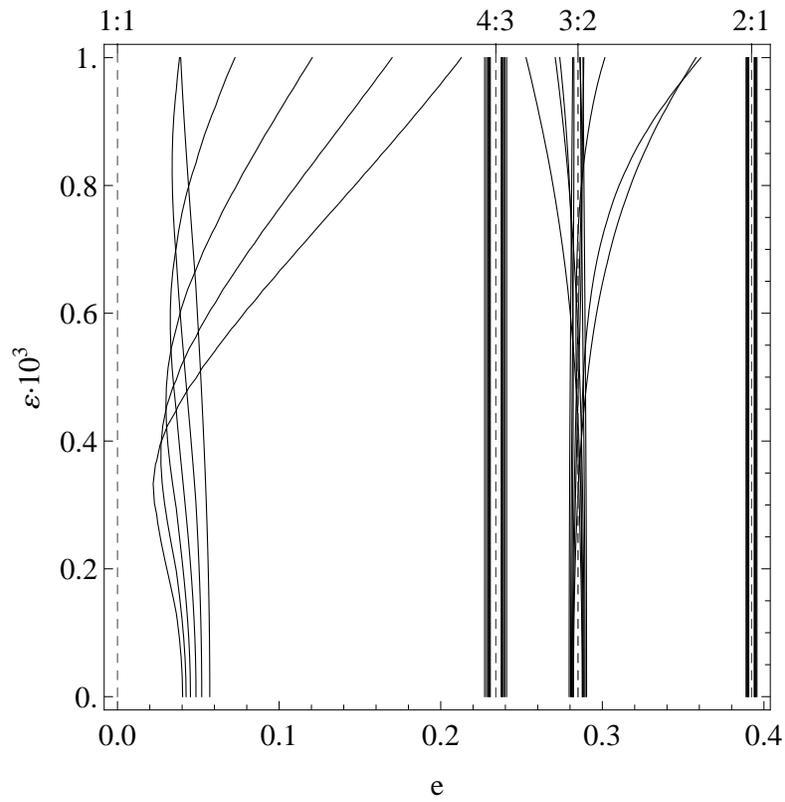}
\caption{Contours obtained from the relation $C(\omega,e,\varepsilon,\mu)$
for different $\omega_k={p\over q}\pm{1\over {k+\gamma}}$ ($\gamma$ being the golden ratio)
with $k=50,\dots,100$, and $\mu=10^{-3}$ in the space
$e\times\varepsilon$ for $\omega_k$ close to the following $p:q$ resonances:
$2:1$, $3:2$, $4:3$, $1:1$.}
\label{f:kam}
\end{figure}

More precisely, the quantity \(\eta\), itself, is given as a function of the
eccentricity $e$ through $\eta=\bar N(e)/\bar L(e)$, as we can notice from the
original equation of motion \equ{MD1}.  Therefore, we are led to introduce the
function
\beq{eq:eta}
C(\omega,e,\varepsilon,\mu)\equiv\bar N(e)/\bar L(e)-
\eta(\omega,e,\varepsilon ,\mu )
\eeq
and we look for contour plots of $C(\omega,e,\varepsilon,\mu)$ in the space
$e\times \varepsilon$ for fixed values of $\omega$, $\mu$. To avoid zero
divisor evaluations we replace $\omega$ with irrational frequencies
$\omega_k$ obtained through the formula
$$
\omega_k=\frac{p}{q}\pm\frac{1}{k+\gamma}  \ ,
$$
where $p/q$ denotes the exact resonant value, $\gamma=(\sqrt{5}-1)/2$,
and $k=50,60,70,80,90,100$. We show the results for the
$2:1$, $3:2$, $4:3$, and $1:1$ resonance in Figure~\ref{f:kam}.
For the 1:1 resonance we can compute the approximations only from above (namely $\omega_k=1+\frac{1}{k+\gamma}$), since the
rotational history of the satellites allows to state that celestial bodies
rotated fast in the past and that they slowed down due to the dissipation,
eventually ending their evolution in a 1:1 resonance, which corresponds to
circular orbits in the case of vanishing tidal torque (see Remark~\ref{ennelle}).
In Figure~\ref{f:kam} the dissipative parameter is fixed to $\mu=10^{-3}$ and
$\varepsilon$ versus $e$ is shown. Since the eccentricity is related to the drift parameter through
$\eta=\bar N(e)/\bar L(e)$, the above picture is essentially equivalent to showing the perturbing versus the drift parameter.

As Figure~\ref{f:kam} shows, using equation \equ{eq:eta}, which depends on the
computation of the parametrization of the invariant attractor, we are able to
relate the shape parameter $\varepsilon$ of the rotating body with the orbital
parameter $e$, thus providing an interesting information from the astronomical
point of view. On the other hand, though typically periodic orbits are used to
approximate invariant tori (just taking the rational approximants to the
irrational - Diophantine - frequency), Figure~\ref{f:kam} provides an
indication of how the invariant KAM tori approximate the main resonances
($2:1$, $3:2$, $4:3$, $1:1$).
In Figure~\ref{f:kam} we observe the accumulation of straight lines, which are
parallel to the dashed lines that indicate the exact positions of the
$p/q=4:3$ and $p/q=2:1$ resonances. Though not being possible to
distinguish the separate lines graphically in Figure~\ref{f:kam}, we confirm that the
curves defined by $C=0$ tend to the dashed curves defined by $\eta(e)=p/q$ for
larger $k$, and arbitrary $\varepsilon$. On the contrary, for the $1:1$ and $3:2$
resonances we observe that the curves start to converge to the dashed lines for small
$\varepsilon$, while they bend for $\varepsilon$ larger. We believe that a higher order
expansion should be computed for larger values of the perturbing parameters,
since getting closer to the resonances the effect of the small divisors becomes
amplified, thus leading to the divergence of the series defining the parametrization.

The $1:1$ resonance is
very common for several natural satellites in the solar system. A future study
of the $1:1$ resonance may provide further information. Beside that, we think
that our relation provides an interesting information concerning how the
oblateness parameter is linked to the eccentricity in the proximity of the
different kinds of resonances.

\section{Conclusions}\label{sec:conclusions}
We have investigated some dynamical features of the dissipative spin--orbit
problem, modeling the dynamics of a non--rigid satellite orbiting on a
Keplerian ellipse around a central planet. In particular we have focused on the
transient time to reach an attractor; this transient time is often decided on
the basis of numerical experiments, integrating the vector field on longer time
scales and looking for the convergence of some quantities providing an
indication that the attractor is reached. Here we have proposed a method which
is based on the analytical development of a dissipative normal form. This
construction allows to compare the normalized frequency with that obtained
integrating the original equations on longer time scales. The normal form,
suitably developed to parametrize invariant attractors, provides also an explicit
expression of the drift term for a specific attractor with fixed frequency.
This allows to give a constraint between the oblateness parameter and the
eccentricity, which may be particularly relevant in astronomical applications.

\vskip.2in

{\bf Acknowledgements.} A.C. was partially supported by PRIN-MIUR
2010JJ4KPA$\_$009, GNFM-INdAM and by the European MC-ITNs Astronet-II and
Stardust. C.L. was partially supported by the ESA-BELSPO grant Prodex C90253
and the Austrian Science Fund (FWF): J3206.

\end{document}